\documentclass[trackchanges,twocolumn]{aastex701}

\begin{document}

\title{Mapping ground-based coronagraphic images to Helioprojective-Cartesian coordinate system by image registration}

\correspondingauthor{Yu Liu}

\author[sname='Sha', gname='Feiyang']{Feiyang Sha}
\affiliation{School of Physical Science and Technology,
	Southwest Jiaotong University, Chengdu 611756, China}
\email[show]{shafy@my.swjtu.edu.cn}
%\author{Yu Liu}

\author[sname='Liu',gname='Yu']{Yu Liu}
\affiliation{School of Physical Science and Technology,
	Southwest Jiaotong University, Chengdu 611756, China}
\email[show]{lyu@swjtu.edu.cn}

\author[sname='Xia',gname='Lidong']{Lidong Xia}
\affiliation{Institute of Space Sciences, Shandong University, Weihai 264209, China}
\email{xld@sdu.edu.cn}

\author[sname='Chen',gname='Yao']{Yao Chen}
\affiliation{School of Physical Science and Technology,
	Southwest Jiaotong University, Chengdu 611756, China}
\affiliation{Institute of Space Sciences, Shandong University, Weihai 264209, China}
\email{yaochen@sdu.edu.cn}

\author[sname='Zhou', gname='Qing']{Qing Zhou}
\affiliation{School of Physical Science and Technology,
	Southwest Jiaotong University, Chengdu 611756, China}
\email{zhouqing@my.swjtu.edu.cn}

\author[sname='Chen', gname='Yangrui']{Yangrui Chen}
\affiliation{School of Physical Science and Technology,
	Southwest Jiaotong University, Chengdu 611756, China}
\email{chenyangrui@my.swjtu.edu.cn}

\author[sname='Zhong', gname='Chuyu']{Chuyu Zhong}
\affiliation{School of Physical Science and Technology,
	Southwest Jiaotong University, Chengdu 611756, China}
\email{zcy20050510@my.swjtu.edu.cn}

\author[sname='Zhang', gname='Xuefei']{Xuefei Zhang}
\affiliation{Yunnan Observatories,
	Chinese Academy of Sciences,
	Kunming 650216, China}
\email{zhangxuefei@ynao.ac.cn}

\author[sname='Song', gname='Tengfei']{Tengfei Song}
\affiliation{Yunnan Observatories,
	Chinese Academy of Sciences,
	Kunming 650216, China}
\email{stf@ynao.ac.cn}

\author[sname='Sun', gname='Mingzhe']{Mingzhe Sun}
\affiliation{Institute of Space Sciences, Shandong University, Weihai 264209, China}
\email{sunmingzhe@sdu.edu.cn}

\author[sname='Yu', gname='Xiaoyu']{Xiaoyu Yu}
\affiliation{Institute of Space Sciences, Shandong University, Weihai 264209, China}
\email{164048770@qq.com}

\author[sname='Li',gname='Haitang']{Haitang Li}
\affiliation{School of Physical Science and Technology,
	Southwest Jiaotong University, Chengdu 611756, China}
\email{lihaitang@swjtu.edu.cn}

\author[sname='Oloketuyi',gname='Jacob']{Jacob Oloketuyi}
\affiliation{School of Physical Science and Technology,
	Southwest Jiaotong University, Chengdu 611756, China}
\affiliation{Department of Physics,
	 Bamidele Olumilua University of Education, 
	 Science and Technology, Ikere-Ekiti 361251, Nigeria}
\email{jacob.oloketuyi@swjtu.edu.cn}

\author[sname='Liu',gname='Qiang']{Qiang Liu}
\affiliation{School of Physical Science and Technology,
	Southwest Jiaotong University, Chengdu 611756, China}
\email{17381352155@163.com}

\author[sname='Wang',gname='Xinjian']{Xinjian Wang}
\affiliation{School of Physical Science and Technology,
	Southwest Jiaotong University, Chengdu 611756, China}
\email{wangxinjian@my.swjtu.edu.cn}
\author[sname='Luo',gname='Qiwang']{Qiwang Luo}
\affiliation{School of Physical Science and Technology,
	Southwest Jiaotong University, Chengdu 611756, China}
\email{lqiwang@my.swjtu.edu.cn}

\author{Xiaobo Li}
\affiliation{Yunnan Observatories,
	Chinese Academy of Sciences,
	Kunming 650216, China}
\email{lixiaobo@ynao.ac.cn}
%\collaboration{all}{The Terra Mater collaboration}

\begin{abstract}

A few ground-based solar coronagraphs have been installed in western China for observing the low-layer corona in recent years. However, determining the Helioprojective Coordinates for the coronagraphic data with high precision is an important but challenging step for further research with other multi-wavelength data. In this paper, we propose an automatic coronal image registration method that combines local statistical correlation and feature point matching to achieve accurate registration between ground-based coronal green-line images and space-based 211 \r{A} images. Then, the accurate field of view information of the coronal green-line images can be derived, allowing the images to be mapped to the Helioprojective Cartesian Coordinates with an accuracy of no less than $0.1''$. This method has been extensively validated using 100 days of coronal data spanning an 11-year period, demonstrating its broad applicability to ground-based coronagraphs equipped with green-line observations. It significantly enhances the scientific value of ground-based coronal data, enabling comprehensive studies of coronal transient activities and facilitating the joint analysis of data from multiple instruments. Additionally, it holds potential for future applications in improving the pointing accuracy of coronagraphs.

\end{abstract}



\section{Introduction} \label{sec:intro}
The solar corona is the outermost layer of the solar atmosphere, extending outward to several solar radii or even farther. 
% There are many intense solar eruptions occurring in the corona, such as flares \citep{Hudson91} and coronal mass ejections (CMEs) \citep{Kahler92}. The high-energy electromagnetic radiation, high-energy particles, and plasma magnetic clouds generated by these solar activities may cause severe space weather effects upon reaching Earth \citep{klein82, Reames99}, affecting people's normal lives. 
The brightness of the corona in the visible light band is extremely low, only about one-millionth as bright as the solar disk \citep{Stix89}, which means that it is typically obscured by strong stray light and not visible. An exception occurs during a total solar eclipse, when the Moon completely covers the solar disk, allowing the corona to be visible. However, such opportunities are rare.
The coronagraph is designed to observe the solar corona during non-total solar eclipse periods. Its fundamental principle involves creating an artificial total solar eclipse by blocking the Sun with an occulter, along with strict stray light suppression \citep{Lyot33}. Utilizing a coronagraph enables continuous monitoring of the solar corona, facilitating a more comprehensive study of its activity and evolution while also allowing for early warnings of hazardous space weather \citep{Schwenn05,Webb12}.

Due to the presence of the occulter, the solar disk is excluded from the coronagraph's field of view (FOV), and the occulter is typically designed to be slightly larger than the solar disk to further suppress stray light. For example, the occulter size is $1.05\ R_\odot$ (solar radius) for \textit{Upgraded Coronal Multichannel Polarimeter} (UCoMP) \citep{Tomczyk19} and \textit{Spectral Imaging CoronaGraph} (SICG) of Shandong University  \citep{sicg, Tang25}, and 1.03 $R_\odot$ for \textit{Yunnan Observatories Green-line Imaging System} (YOGIS) \citep{Ichimoto99, Liu14}. 
%加上关于半径的描述
It is important to note that the occulter does not consistently align with the Sun due to various factors, including limited accuracy in pointing and guiding, spar flexure, wind conditions, and so on. 
Additionally, due to the telescope's varying force states in different orientations, together with temperature fluctuations of the optical components, the detector does not consistently align with the focal plane, complicating the precise determination of the solar radius in the image. 
Consequently, without a clear indicator like the solar limb, determining the solar center and solar radius in the observed image becomes challenging, which has been mentioned in many previous works (e.g., \citealp{Bak16}, \citealp{Mettes22}, \citealp{West23}). For example, when studying transient coronal activities (e.g., \citealp{Tian12,Shen22}) or combining data from multiple instruments (e.g., \citealp{Fletcher11}). 
% First, when using data from a single coronagraph to study transient coronal activities, such as CMEs, flows, oscillations, and waves (e.g., \citealp{Tian12,Shen22}), it is crucial to align observational data over time to obtain the differential or time-distance diagram.
% Second, when combining coronagraphic observations with other wavelengths or instruments (e.g., \citealp{Fletcher11}), it is essential to determine the absolute position of the solar disk and map these different images to a common coordinate system, such as the Helioprojective Cartesian Coordinates \citep{Thompson06}, which is also essential for routine measurements of the coronal magnetic field using coronagraphs. \citep{Liu08,Yang24}.

The solar center in coronagraph images must be obtained using other references \citep{Morrill06}. For space-based coronagraphs, such as SOHO/LASCO \citep{lasco}, it is determined by the positions of stars within the FOV, the time of observation, and the spacecraft's roll angle \citep{Morrill06}. In contrast, these cannot be applied to ground-based coronagraphs due to atmospheric scattering and refraction. Some coronagraphs are equipped with an off-axis guiding scope to observe the solar disk, from which the solar center is obtained. However, the small difference in pointing between the coronagraph and the guiding scope does cause errors.

Similar challenges exist in high-resolution solar observations. Previous studies employed image registration methods to align high-resolution solar images with reference images with known FOV, such as full-disk solar images, thereby indirectly obtaining the absolute positions of the images (e.g., \citealp{Feng18, Ji19}). However, this method has not yet been applied to coronal images, partly due to their low signal-to-noise ratios (SNR) and the challenge in finding corresponding reference images. 
Fortunately, \cite{Zhang22} compared the coronal green-line (Fe XIV, 5303 \r{A}) intensities with the extreme-ultraviolet (EUV) images of SDO/AIA \citep{aia} and found that the intensity distributions of the 211 \r{A} images exhibit very high correlation coefficients with the coronal green-line, ranging from 0.89 to 0.99. Thus, the 211 \r{A} images can serve as ideal reference images for the registrations of coronal green-line images.

Image registration is a process of aligning different images by applying similarity transformations involving translation, rotation, and scaling, if there is no distortion \citep{Zitova03}.
The main challenge lies in accurately identifying the parameters for each type of transformation.
Generally, image registration methods can be categorized into two main types, i.e, feature-based and area-based. 

Feature-based methods align images by extracting and matching salient features in the images, such as edges and points with high-frequency information.
However, this method is difficult to apply to coronal images, because there are almost no clear edges except for the occulter edge, and the low-frequency components dominate, making it challenging to extract features.

Area-based methods, also known as correlation-like methods, compare the intensity distribution of regions within the images using cross-correlation (CC) or phase-correlation (PC) to provide translation parameters but without rotation and scaling parameters \citep{Pratt74}. 
Some studies align coronal images observed during total solar eclipses based on area-based methods. \cite{Druckmuller09} applied a modified phase correlation method to coronal images, which is capable of estimating all three transformation parameters. However, it cannot be automated due to the need for manually adjusting the initial parameters to achieve optimal alignment results. \cite{Liang21} combine CC with blind deconvolution and the noise-adaptive fuzzy equalization method to enhance the features of a small coronal region, enabling precise translation transformations. However, rotation and scaling are not included \citep{Liang21}, which are essential for the registration between images captured by different devices. This is because of the uncertainty of the solar radius and the polar angle.

The two methods mentioned above were developed for total solar eclipse coronal images, which may not be suitable for coronagraph images, characterized by lower image quality and SNR due to atmospheric scattering of the solar disk. \cite{Feng18} proposed a registration method that combines local statistical correlation and feature-matching by dividing the image into many smaller subregions, and calculating the cross-correlation of the small areas to obtain matching points. This method is very accurate. However, it is limited to solar disk images. Moreover, it requires appropriate initial FOV parameters, which can be challenging in practical applications.

Based on the work of \cite{Feng18}, we develop an Automatic and high-Precision RegIstration aLgorithm (APRIL) that achieves an automatic registration between coronal green-line and SDO/AIA 211 \r{A} images, leading to the mapping of the green-line images to the Helioprojective Cartesian Coordinates (HCC).
%% The "ht!" tells LaTeX to put the figure "here" first, at the "top" next
%% and to override the normal way of calculating a float position.
%% The asterisk after "figure" tells the compiler to span multiple columns
%% if a two column style is selected.
\section{Data and Method}
\subsection{Data}
The coronal green-line images used for evaluating the APRIL method were captured by YOGIS, the first routinely operated coronagraph in China \citep{Liu14}. 
It has an aperture of 10 cm and a focal length of 1490 mm, capable of imaging the coronal green line at 5303 \r{A} with a bandwidth of 1 \r{A} \citep{Ichimoto99}. 
YOGIS has been operating continuously at Lijiang Observatory since October 2013, providing 11 years of coronal green-line observations.
%We utilized coronal data observed by YOGIS across two years, 2015 and 2024. Specifically, one or two days of coronal data were selected from each month, resulting in a total of 34 days.
In this study, a total of 100 days of coronal green-line data were used, spanning the 11 years from 2013 to 2024, with one or two days selected from each month.

Furthermore, to comprehensively assess the performance of the APRIL method on different coronagraphs, one-day data captured by SICG on May 11, 2024, was also selected. SICG is a 20-cm aperture coronagraph of the Chinese Meridian Project Phase II \citep{wang24}, capable of imaging the coronal green line and red line \citep{sicg,Tang25}.

The corresponding SDO/AIA 211 \r{A} EUV images were obtained for the relevant time using the Python package hvpy\footnote{\url{https://hvpy.readthedocs.io/}}, which provides researchers with a convenient way to access solar and heliospheric datasets. 

In the following subsection, we illustrate the steps of the APRIL method using coronal data obtained from YOGIS on November 11, 2024, at 07:38:34 UT as a representative example.
This coronal green-line image and its corresponding SDO/AIA 211 \r{A} image are shown in Figure \ref{fig:step0}(a) and (b). 
\subsection{Method}

\subsubsection*{Step 1: Preliminary Registration}
\begin{figure*}[ht!]
\plotone{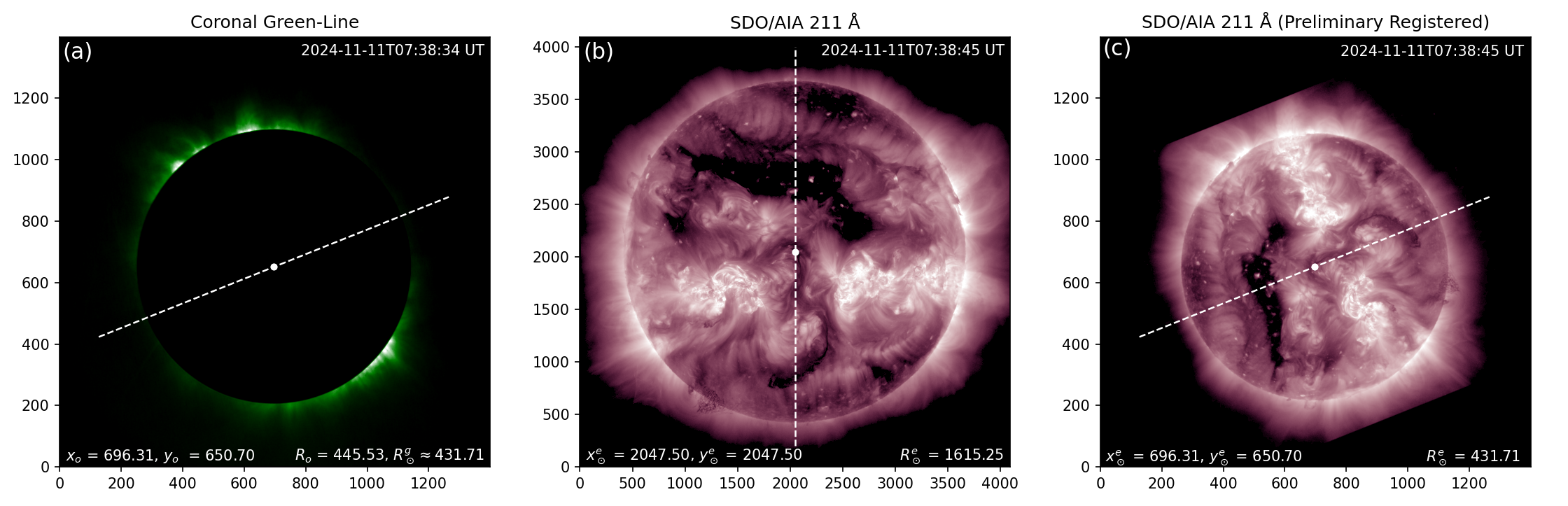}
\caption{Panels (a) and (b) are the raw images of coronal green-line and SDO/AIA 211 \r{A} observed on November 11, 2024, at 07:38:34 UT and 07:38:45 UT. Panel (c) is transformed from panel (b) and preliminary registered with panel (a).
The center coordinates and radius of the solar disk or the occulter are given at the bottom of each panel. The solar center and polar axis are marked by a circle and a dashed line, respectively, but are not accurate in panel (a).\label{fig:step0}}
\end{figure*}

Before precise registration, the first step is to roughly align the green-line and 211 \r{A} EUV images by applying a similarity transformation to the latter.

The pixel size is $1400\times1400$ for Figure \ref{fig:step0}(a) and $4096\times4096$ for Figure \ref{fig:step0}(b). The pixel resolutions of the two images are about 2 and 0.6 arcsec/pixel. If the two images are aligned, they should have the same pixel resolution and solar radius.
Therefore, the latter is downsampled by a factor of $k_0$, which satisfies:
\begin{equation}
	\label{eq:k0}
	% k_0=\frac{R_\odot^g}{R_\odot^e}%\approx \frac{1}{3.74},
	k_0=R_\odot^g/R_\odot^e,
\end{equation}
where $R_\odot^g$ and $R_\odot^e$ are the solar radii of panels (a) and (b). Note that the solar radius is not precisely fixed and varies throughout the year. For YOGIS, we have ten occulters of different sizes to accommodate these variations in the solar radius. The ratio between their sizes is maintained at approximately 1.03, although some deviations exist. So, we have
\begin{equation}
	\label{eq:approx}
	\begin{array}{cc}
		(x_\odot^g,y_\odot^g) \approx (x_o,y_o)
		% =(696.27,650.72)   
		,\quad R_\odot^g \approx R_{o}/1.03
		% =431.53
		,
	\end{array}
\end{equation}
where $(x_\odot^g, y_\odot^g)$ and $(x_o, y_o)$ are the center coordinates of the solar disk and occulter, respectively.
The $R_o$ is the occulter's radius, which is derived by performing a circle fit on its edge. 
These values are shown in Figure \ref{fig:step0}(a). In this step, we temporarily ignore the deviations and assume them to be equal. In this case, $k_0$ is calculated to be $0.267$. 
Then, the pixel size of panel (b) is adjusted to match that of panel (a), $1400\times1400$, and panel (b) is translated so that the solar centers of both images are aligned, even if not precisely.

After scaling and translation, the last operation is rotation. By converting the images from Cartesian to polar coordinates, the rotational difference between them is transformed into a translational shift along the angular axis. The rotation angle $\theta_0$ can then be determined by computing the CC between the transformed images and measuring the peak displacement along the angular axis. This peak location directly corresponds to the rotation angle in the original Cartesian coordinate system. In this case, it is $111.9^\circ$ (counterclockwise).

The resulting image is shown in Figure \ref{fig:step0}(c). Note that the similarity transformations applied above are inaccurate, and the errors arise from Equation \ref{eq:approx} and the rotation angle. 

\subsubsection*{Step 2: Accurate Registration}
In this step, we apply a similarity transformation on Figure \ref{fig:step0}(c) while keeping Figure \ref{fig:step0}(a) fixed to achieve an accurate registration between the two images including translation, scaling, and rotation. The resulting transformation parameters can then be used to deduce the precise solar center coordinate, solar radius, and polar axis direction of the green-line image, as depicted in Figure \ref{fig:step0}(a).

\begin{figure*}[ht!]
	\plotone{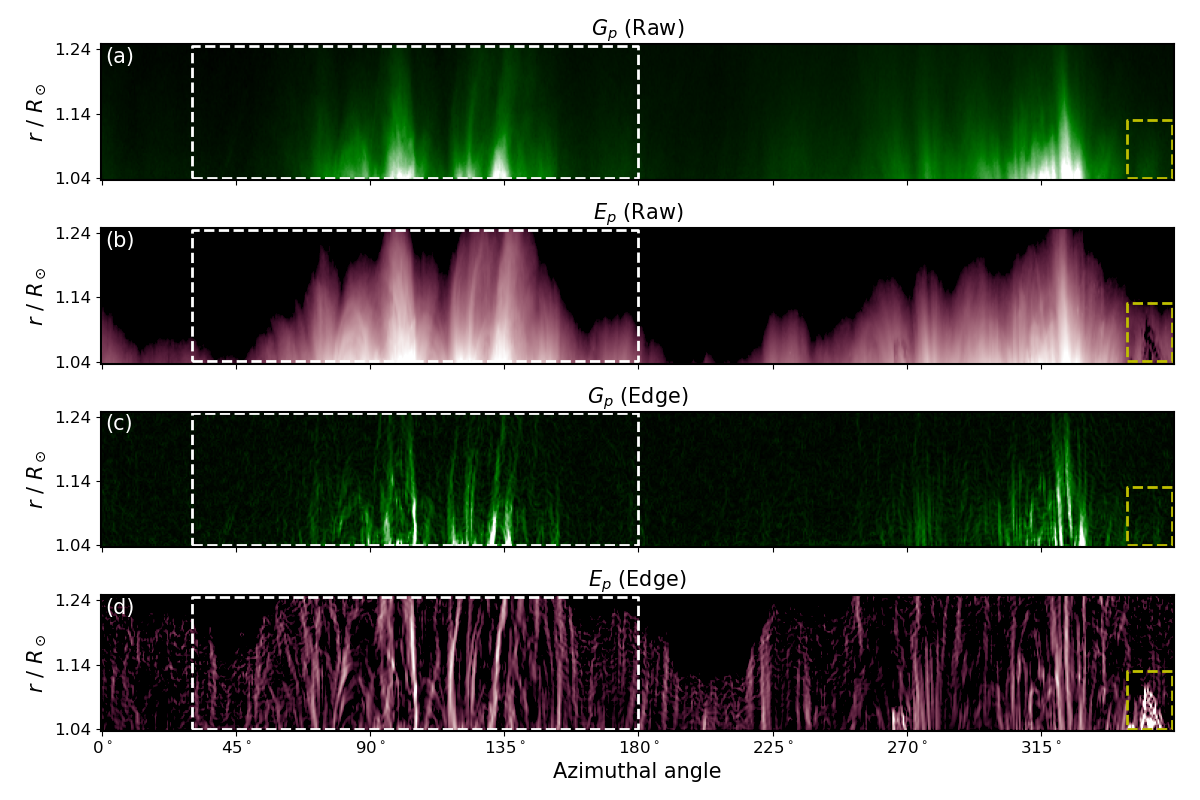}
	\caption{The upper two panels are the polar transformed images of $G$ and $E$, while the lower two panels present their edge magnitudes. All the panels have the same size, $91 \times 720$, covering a radial range from 1.04 $R_\odot$ to 1.25 $R_\odot$ with a full $360^\circ$ azimuthal range, with $0^\circ$ or $360^\circ$ corresponding to the right side and angles increase counterclockwise (90° top, 180° left, 270° bottom, completing the 360° cycle at the right limb). The regions enclosed by the dashed white boxes are examples of a pair of subregions, starting from $30^\circ$ and spanning an azimuthal width of $150^\circ$. A prominence is seen in the dashed yellow box in panels (b) and (d), but absent in panels (a) and (c).
	\label{fig:polar}}
\end{figure*}

For simplicity, Figure \ref{fig:step0}(a) and Figure \ref{fig:step0}(c) are denoted as $G$ and $E$, respectively. $G$ and $E$ are transformed from Cartesian coordinates to polar coordinates. The regions extending approximately 4 pixels (about 0.01 $R_\odot$) beyond the occulter contain diffraction rings that can interfere with the analysis of coronal structures, while the upper regions ranging from $1.04\ R_\odot$ to $1.25\ R_\odot$ are retained, denoted as $G_p$ and $E_p$ (Figure \ref{fig:polar}(a) and (b)). Furthermore, the regions beyond 1.25 $R_\odot$ are also excluded from analysis due to incomplete SDO/AIA data coverage resulting from the instrument's square field-of-view limitation.

Then, the edge magnitudes of $G_p$ and $E_p$, illustrated in Figure \ref{fig:polar}(c) and (d), are computed using the Scharr operator \citep{Weickert02}, which has a symmetric kernel and provides better rotational invariance compared to other edge enhancement operators. This step, referred to as ``Edge Enhancement", is crucial. In the following sections, we will conduct a comparative analysis of the results obtained with and without the ``Edge Enhancement" step.
The poles of the polar coordinates correspond to the same coordinate $(x_o,y_o)$, ensuring consistency between $G_p$ and $E_p$ and the absence of the occulter in $G_p$. However, due to the deviation in Equation \ref{eq:approx}, $(x_o,y_o)$ does not correspond to the same absolute location between $E$ and $G$, resulting in a distortion between $E_p$ and $G_p$. Consequently, direct alignment through similarity transformation is not feasible.

\begin{figure*}[ht!]
\plotone{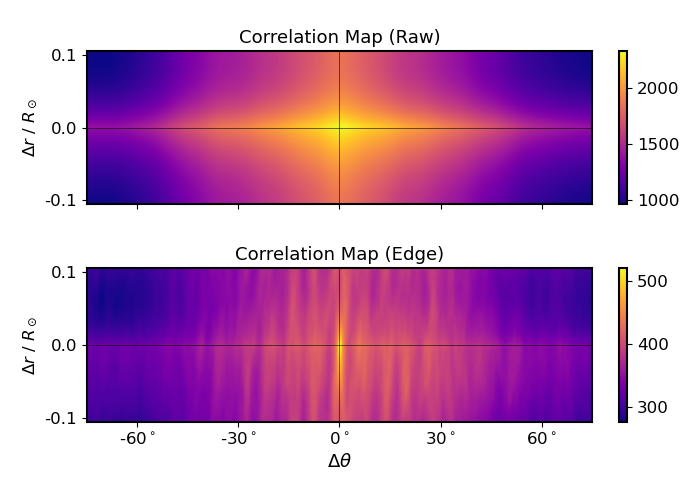}
\caption{The correlation maps calculated by Equation \ref{eq:cor}. The upper and lower panels show the correlation map without and with the ``Edge Enhancement" step, respectively. Each pixel value corresponds to the cross-correlation of $G_p^i$ and $E_p^i$, where $E_p^i$ is shifted by $(\Delta \theta, \Delta r)$.\label{fig:cor}}
\end{figure*}

By segmenting the images, the distortion can be decomposed into multiple shifts, each corresponding to a pair of subregions. 
The subregions are segmented along the azimuthal axis (see the dashed white boxes on the lower two panels of Figure \ref{fig:polar}), with the $i_{\text{th}}$ ($i\in[1,36])$ subregion covering an azimuthal range of $[10i,10i +w]$ degrees, where $w$ is the azimuthal width. Adjacent subregions partially overlap to enhance data utilization.
The subregions extracted from $G_p$ and $E_p$ are denoted as $G_p^i$ and $E_p^i$, respectively. The shift between them is derived using CC as follows:
\begin{equation}
	\label{eq:cor}
	\begin{array}{cc}
		Cor^i= \mathcal{F}^{-1} \left\{ \mathcal{F} \{ G_p^i \} \cdot \mathcal{F}^* \{ E_p^i\} \right\},
	\end{array}
\end{equation}
where $\mathcal{F}$ and $\mathcal{F}^{-1}$ denote the Fourier transform and its inverse, respectively. The superscript $^*$ indicates the complex conjugate. The resulting correlation map, $Cor^{i}$, is illustrated in Figure \ref{fig:cor}. The offset between $G_p^i$ and $E_p^i$ is derived from the coordinates of the maximum pixel value of the correlation map, $(\Delta\theta^i,\Delta r^i)$, and is refined using a localized upsampled discrete Fourier transform, achieving subpixel precision \citep{Guizar-Sicairos:08}. Panel (b) applies the ``Edge Enhancement" step while panel (a) does not. Clearly, panel (b) exhibits sharper and more concentrated peaks near the maximum point. This demonstrates that the edge filter enhances the accuracy and reliability of the correlation analysis.

Taking $(\theta_g^i,r_g^i)$, the center of $G_p^i$, as the reference point, the corresponding point in $E_p^i$ is derived by:
\begin{equation}
	\label{eq:point}
	\begin{array}{cc}
		(\theta_e^i,r_e^i)=(\theta_g^i,r_g^i)-(\Delta\theta^i,\Delta r^i).
	\end{array}
\end{equation}
By converting these two points back to the original Cartesian coordinates, a pair of matching points for the registration between $G$ and $E$ is obtained, denoted as $(x_g^i,y_g^i)$ and $(x_e^i,y_e^i)$, respectively. Thus, 36 pairs of matching points are derived from 36 pairs of subregions. These matching points are subsequently used to compute the similarity transformation parameters for image $E$.

Note that the accuracy of CC is influenced by the azimuthal width $w$. A smaller $w$ reduces the information content in $G_p^i$ and $E_p^i$, while a larger $w$ results in significant image distortion. Both cases degrade CC performance, and the optimal value of $w$ is found to be $150^\circ$. 

Ideally, each pair of matching points satisfies:
\begin{equation}
	\label{eq:matrix}
	\left(\begin{array}{c}
		x_{e}^{i} \\
		y_{e}^{i}
	\end{array}\right)=\left[\begin{array}{cccc}
		1 & 0 & x_{g}^{i} & -y_{g}^{i} \\
		0 & 1 & y_{g}^{i} & x_{g}^{i}
	\end{array}\right]\left(\begin{array}{c}
		\Delta x \\
		\Delta y \\
		k c \\
		k s
	\end{array}\right),
\end{equation}
where $\Delta x$ and $\Delta y$ represent the translation parameters, $kc=k\cos\theta$ and $ks=k\sin\theta$, with $k$ and $\theta$ representing the scaling and rotation parameters. These two equations are insufficient to determine the 4 unknown parameters in Equation \ref{eq:matrix}. Thus, additional matching points are required to determine the parameters. However, the results derived from Equation \ref{eq:cor} may contain errors due to several factors. 
For instance, errors may arise from insufficient coronal features in $G_p^i$ (e.g., polar regions), observational interference such as airborne dust in the FOV, and differences between the coronal green-line and 211 \r{A} EUV emissions, as illustrated by the prominence in the yellow boxes of Figure \ref{fig:polar}.

The random sample consensus (RANSAC) method is employed to automatically exclude erroneous matching points. First, $n$ pairs of matching points are randomly selected (we set $n=25$) and used to extend Equation \ref{eq:matrix} as follows:
\begin{equation}
	\left(\begin{array}{c}
		x^{1}_{g} \\
		y^{1}_{g} \\
		x^{2}_{g} \\
		y^{2}_{g} \\
		\vdots \\
		x^{n}_{g} \\
		y^{n}_{g}
	\end{array}\right)=\left[\begin{array}{cccc}
		1 & 0 & x^{1}_{e} & -y^{1}_{e} \\
		0 & 1 & y^{1}_{e} &  x^{1}_{e} \\
		1 & 0 & x^{2}_{e} & -y^{2}_{e} \\
		0 & 1 & y^{2}_{e} &  x^{2}_{e} \\
		\vdots & \vdots & \vdots & \vdots \\
		1 & 0 & x^{n}_{e} & -y^{n}_{e} \\
		0 & 1 & y^{n}_{e} &  x^{n}_{e} 
	\end{array}\right]\left(\begin{array}{c}
		\Delta x \\
		\Delta y \\
		k c \\
		k s
	\end{array}\right),
\end{equation}
which simplifies to $\mathbf{V}=\mathbf{M}\mathbf{X}$, where $\mathbf{X}$ is calculated using the least squares method as follows: $$\mathbf{X}=\mathbf{(M^T M)^{-1}}\mathbf{M^T V},$$ 
from which the similarity transformation parameters $(\Delta x,\Delta y),\ k$ and $\theta$ are derived.
Next, the unselected $(36-n)$ pairs of matching points are used to check the outcomes, evaluated by the median distance between $(x_g^j,y_g^j)$ and $(\hat{x}_g^j,\hat{y}_g^j)$, where $j$ corresponds to the unselected points, and $(\hat{x}_g^j,\hat{y}_g^j)$ is calculated as follows:
\begin{equation}
	\label{eq:matrix2}
	\left(\begin{array}{c}
		\hat{x}_g^j \\
		\hat{y}_g^j \\
		1
	\end{array}\right)=\left[\begin{array}{ccc}
		k \cos \theta & -k \sin \theta & \Delta x \\
		k \sin \theta & k \cos \theta & \Delta y \\
		0 & 0 & 1
	\end{array}\right]\left(\begin{array}{c}
		x_{e}^{j} \\
		y_{e}^{j} \\
		1
	\end{array}\right).
\end{equation}
These processes, including random selection and validation, are repeated 100 times, and the solution $\mathbf{X}$ yielding the minimum median distance is identified as optimal and retained.

The similarity transformation parameters derived from $\mathbf{X}$ are used to transform $E$. The transformed image $E'$ resembles $G$ more closely but is not yet fully aligned. 
This discrepancy arises from inaccuracies in the solar radii and the differences in the absolute positions of $(x_o,y_o)$ between $G$ and $E$.
Therefore, we set
\begin{equation}
	E=E',
\end{equation}
repeat the accurate registration to generate a new $E'$, and iterate the process.
During the iteration, only $E$ is updated, while other parameters, such as the solar radii and poles of the polar coordinates for both images, remain unchanged. 
When $(\Delta x, \Delta y)$ and $\theta$ converge to 0, $G$ and $E$ are considered fully registered, and the iterations terminate.

The registration process can be summarized as follows:
\begin{enumerate}
	\item Perform a preliminary registration between the green-line and 211 \r{A} EUV images.
	\item Transform the images into polar coordinates and apply the Scharr operator.
	\item Segment the images into subregions and compute the shift between each pair of subregions using CC.
	\item Apply the RANSAC method to automatically eliminate erroneous shifts and derive the similarity transformation parameters.
	\item Use the above parameters to transform the EUV image.
	\item If the similarity transformation parameters converge, terminate the process. Otherwise, update the EUV image and repeat the process from Step 2.
\end{enumerate}

\section{Results}
\subsection{One-day Results}
\begin{figure*}[ht!]
\plotone{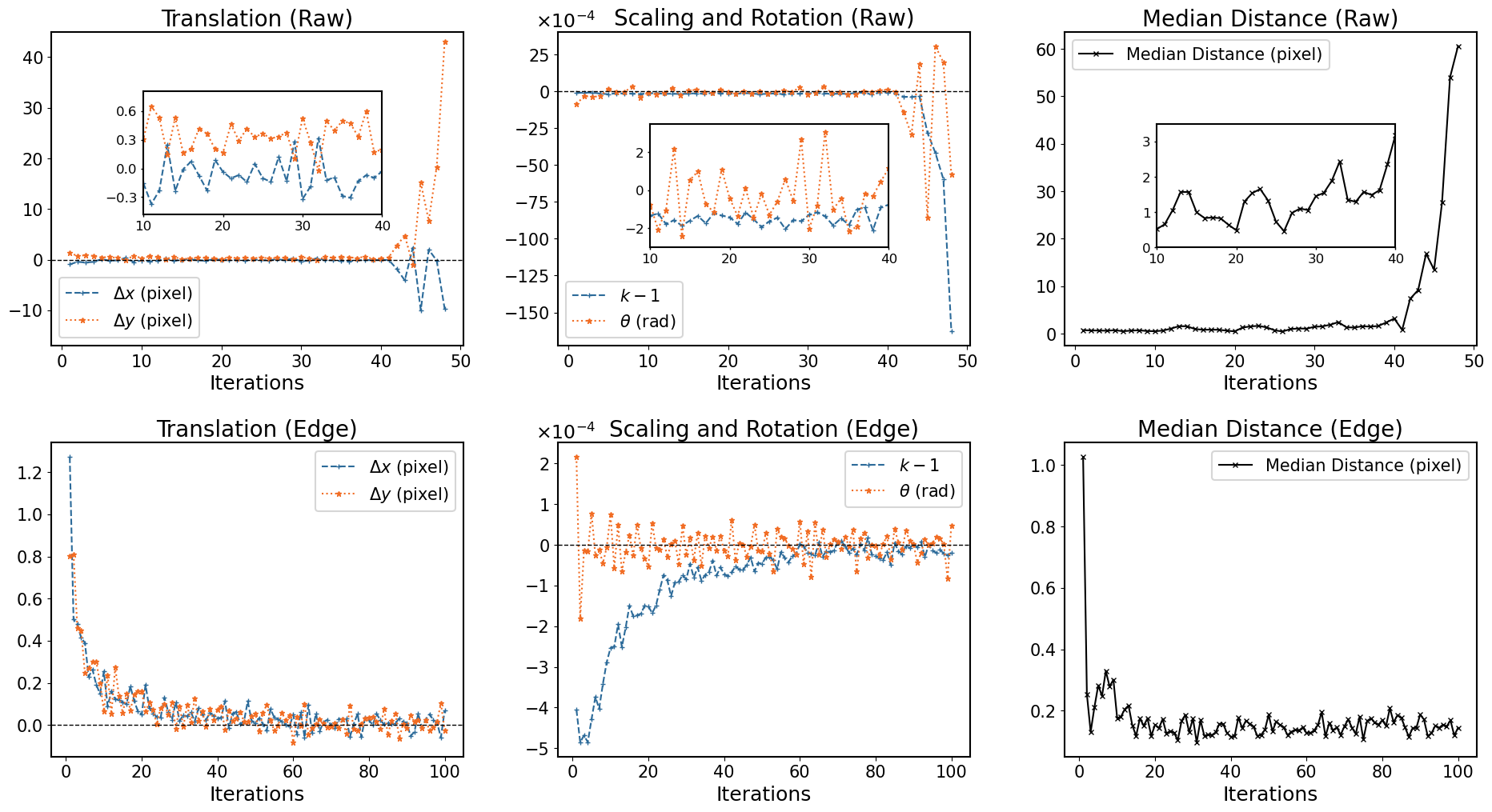}  
\caption{The changes in $\Delta x$, $\Delta y$, $k$, $\theta$, and the median distance between $(x_g^j,y_g^j)$ and $(\hat{x}_g^j,\hat{y}_g^j)$ over iterations. The upper and lower panels show the parameters computed without and with the ``Edge Enhancement" step, respectively. The subpanels within each upper panels provide magnified views, covering the range from 10 to 40 iterations.\label{fig:iteration}}
\end{figure*}

Figure \ref{fig:iteration} illustrates the evolution of the similarity transformation parameters and the errors (median distance between $(x_g^j,y_g^j)$ and $(\hat{x}_g^j,\hat{y}_g^j)$) over the iterations. The upper panels show the parameters computed without the ``Edge Enhancement" step. These parameter values do not converge during the iteration process. Instead, they initially remain stable, but later diverge rapidly, leading to convergence failure. This may arise from the significant uncertainty in identifying the maximum point in the correlation map (Figure \ref{fig:cor} (a)).
This has been verified using a wide range of coronal datasets, showing similar results. In contrast, the lower panels illustrate the parameters computed with the ``Edge Enhancement" step.
As the iterations progress, $(\Delta x, \Delta y)$, $\theta$ and errors gradually converge to 0, and $k$ approaches 1. The process typically converges within 50 iterations, with the error dropping below 0.1 pixel at convergence. 
Based on test results, the iteration terminates when $|\Delta x| < 0.05$ pixel and $|\Delta y| < 0.05$ pixel, corresponding to approximately $0.1''$.

\begin{figure*}
\plotone{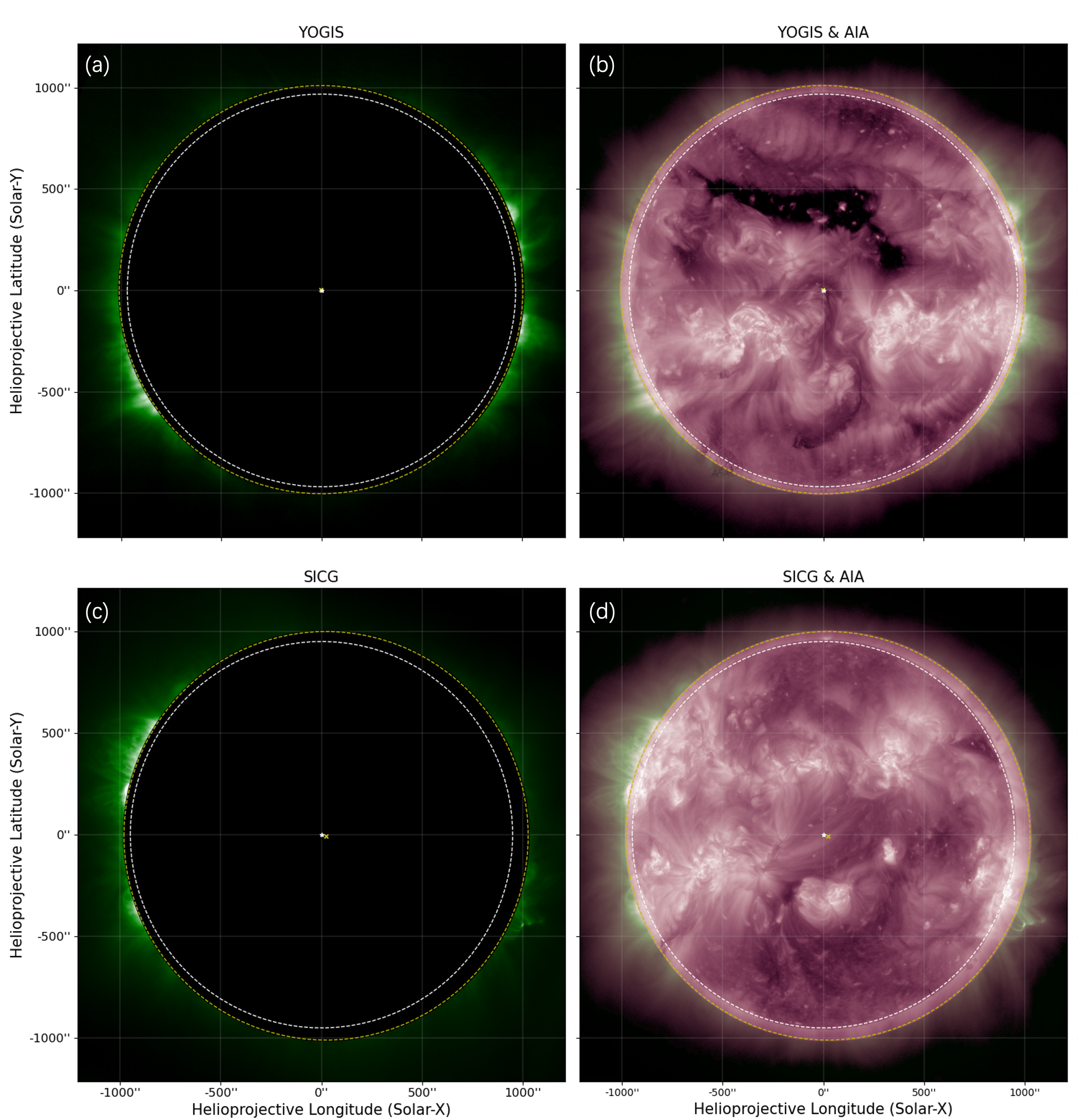}
\caption{Panel (a) and (b) are the coronal green-line image of YOGIS registered using the APRIL method and its composite image with SDO/AIA 211 \r{A} image. Panel (c) and (d) are those observed by SICG on May 11, 2024.
	All panels are mapped to the HCC. The white and yellow dashed circles illustrate the solar limbs and occulter edges, respectively. The white asterisks and yellow plus signs represent the centers of the Sun and the occulter. (An animation of panel (a) and (b) is available online. The additional panel (a') overlays a mask on panel (a) to achieve a better visual effect.) \label{fig:registered_img}}
\end{figure*}

Initially, the precise solar center coordinate of Figure \ref{fig:step0}(c), $(x^e_\odot,y^e_\odot)$, are set to $(x_o, y_o)$, whereas those of Figure \ref{fig:step0}(a), $(x^g_\odot,y^g_\odot)$, are unknown. After applying the similarity transformations to Figure \ref{fig:step0}(d), the solar center coordinates of $G$ and $E'$ are aligned, i.e., $(x^g_\odot,y^g_\odot)=(x_\odot^{e'},x_\odot^{e'})$, where the latter satisfies:
\begin{equation}
	\label{eq:matrix1}
	\left(\begin{array}{c}
		x_\odot^{e'} \\
		y_\odot^{e'}\\
		1
	\end{array}\right)=
	\left[\begin{array}{ccc}
		K \cos \Theta & -K \sin \Theta & \Delta X \\
		K \sin \Theta & K \cos \Theta & \Delta Y \\
		0 & 0 & 1
	\end{array}\right]
	\left(\begin{array}{c}
		x_{\odot}^{e} \\
		y_{\odot}^{e} \\
		1
	\end{array}\right),
\end{equation}
where $(\Delta X,\Delta Y),\ K$ and $\Theta$ represent the similarity transformation parameters for the entire iteration process, obtained from the cumulative product of rotation matrices at each iteration (the matrix in Equation \ref{eq:matrix2}). 
The solar radius of $G$ is given by $r_\odot^g=Kr_\odot^e$, and the polar angle is $\theta_0+\Theta$. Using these parameters, the coronal green-line images can be mapped to the HCC, as illustrated in Figure \ref{fig:registered_img}(a) and (b).
Clearly, the coronal features in the green-line and 211 \r{A} EUV images align perfectly, demonstrating the effectiveness of our registration method. 

\begin{figure*}
	\plotone{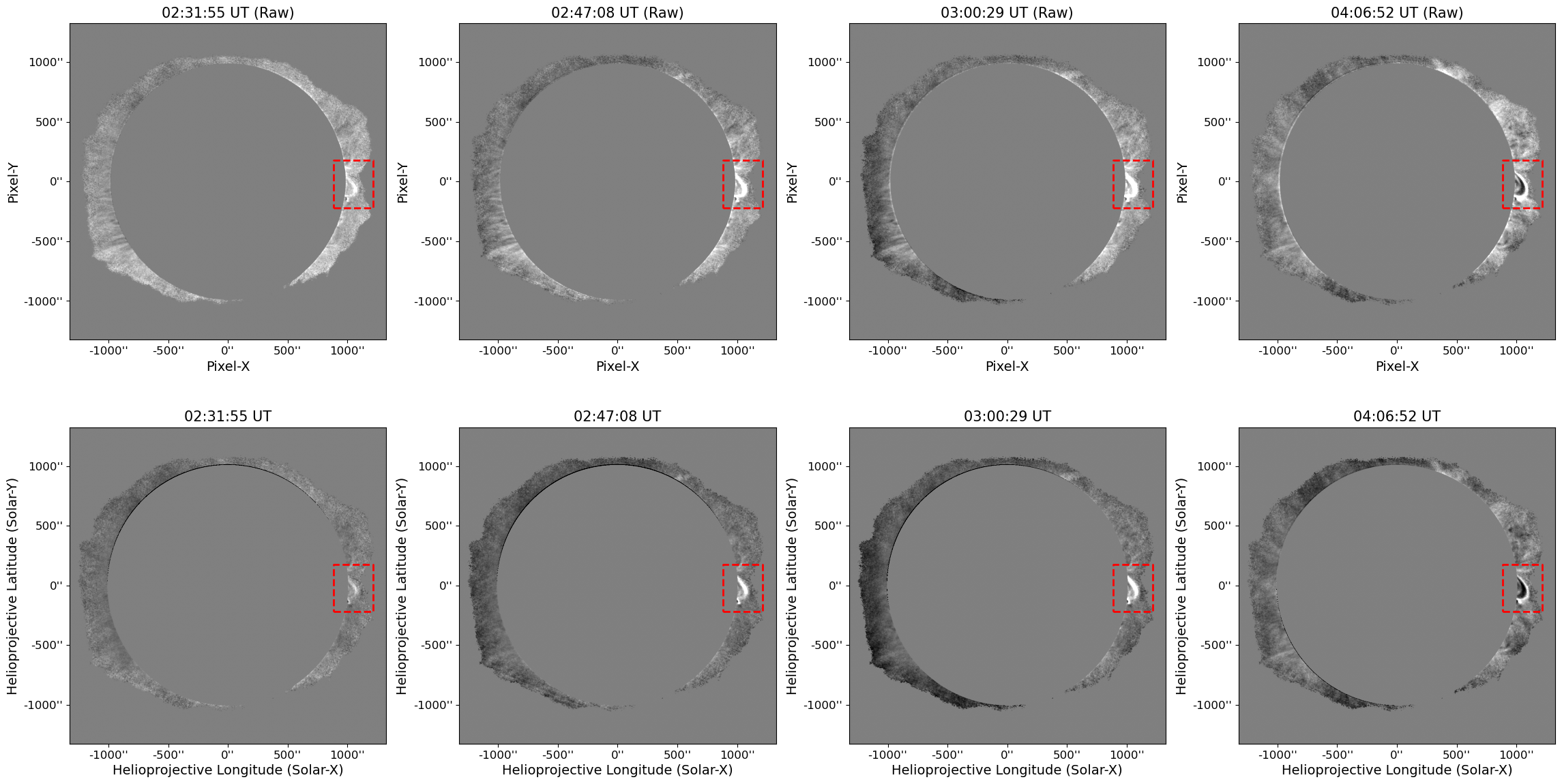}
	\caption{YOGIS coronal green-line base-difference images. All panels were observed on November 11, 2024. The base image was taken at 02:07:21 UT. The top panels show the raw images co-aligned by the occulter, while the bottom panels show images mapped to the HCC using the APRIL method. An erupting loop is visible in the dashed red box of each panel.\label{fig:difference}}
\end{figure*}

We applied the APRIL method to all the coronal green-line images observed by YOGIS on November 11, 2024, and combined them into an animation (see Figure \ref{fig:registered_img} animation). The animation demonstrates that the coronal structure remains stable as the occulter moves over time.
Additionally, Figure \ref{fig:difference} presents the base-difference images of the coronal green-line, with the top panels displaying raw images co-aligned by the occulter and the bottom panels showing the images registered by our method. 
The top panels reveal significant coronal differences on the solar west side, including an apparent loop structure in the dashed red box on the right, caused by a combination of coronal activities and image misalignment. In contrast, the bottom panels exhibit no significant differences, with weaker loop structures compared to the top panels, indicating only coronal activities. This clearly demonstrates the effectiveness and high precision of the registration results.

To validate the effectiveness of the APRIL method for other coronagraphs, we also applied it to coronal green-line images observed by SICG on May 11, 2024. Compared to YOGIS, the only parameter that requires adjustment is the ratio between the occulter radius and the solar radius, as described in Equation \ref{eq:approx}. The ratio is approximately 1.03 for YOGIS and 1.05 for SICG. All other steps remain unchanged. The results, displayed in Figure \ref{fig:registered_img}(c) and (d), demonstrate high performance and accuracy. Thus, we confirm that the APRIL method can be widely applied to all coronagraphs used for coronal green-line observations.

\subsection{Statistical Results}
A total of 34 days of coronal images selected from 2015 and 2024 are used to evaluate the precision of the APRIL method, with 1 or 2 days chosen from each month (see the first column of Table \ref{tb1}). For some months, only one day of data was selected, or no data at all, typically due to the inability to conduct observations during the rainy season in Lijiang in the summer. 

The coronal green-line data observed by YOGIS in 2015 and 2024 exhibit notable differences in image quality. For this reason, we selected data from these two years to demonstrate the effectiveness of our method under varying data quality conditions.

In 2015, the coronagraph had just been moved to Lijiang for about a year, and the observation building had not been constructed. Coronal observations were conducted in a temporary dome in the wild. At that time, the temporary observation assistant had limited experience and training, which resulted in issues such as inaccurate focusing, untimely tracking of the sun, and sustained accumulation of dust on the objective lens without cleaning \citep{Sha23,sicg}. As a consequence, there were only a few dozen images or fewer each day, and the data quality was low. 

In February 2023, the YOGIS observation system was significantly upgraded, including the replacement of the camera and improvements to the data storage system, pointing system, and observation software \citep{song25}, leading to substantial enhancements in data quality and observational efficiency. From then on, the data typically amount to several hundred or even a thousand images per day. Thus, the coronal data from 2024, acquired during solar maximum, are of the highest quality since 2013.

To save testing time, we uniformly selected one-third of the daily data from 2024 as the test sample. The sampled and total data counts are listed in the second column of Table \ref{tb1}.

The RANSAC algorithm employed in the APRIL method involves stochastic processes. We applied it four times to each coronal image. In most cases, the iterative process successfully converges, and the corresponding convergence rates are given in the fourth column of Table \ref{tb1}.
For data that converged in all four repetitions, the standard deviation of the solar center coordinates, solar radius and polar angle were obtained, 
and their mean values in each day are listed in table \ref{tb1}, named as $(\overline{X_{std}}, \overline{Y_{std}})$, $\overline{R_{std}}$ and $\overline{\theta_{std}}$. The average solar radius of each day, $\overline{R_\odot}$ is also presented in the table. These values closely match the actual case, with the solar radius being smaller during summer near aphelion and larger during winter.

To better compare the registration results among these days, we define the rank for each day as follows:
\begin{eqnarray}
	\label{eq:rank}
	Rank = &(100\times\text{Convergence Rate}+ \nonumber \\
	&Rank_x+Rank_y+Rank_r)/4.
\end{eqnarray}
It is essentially an average of four components. The first component corresponds to the Convergence Rate, with higher values leading to higher scores, capped at a maximum of 100. The other components correspond to $\overline{X_{std}}$, $\overline{Y_{std}}$, and $\overline{R_{std}}$, and they follow similar definitions. For instance, $Rank_x$ is defined as:
\begin{equation}
	Rank_x=[1-\frac{\overline{X_{std}}/\overline{R_\odot}}{\text{max}(\overline{X_{std}}/\overline{R_\odot})}]\times100.
\end{equation}
The polar angle $\theta$ is not included in Equation \ref{eq:rank} because its value is significantly smaller than that of other parameters.

The rank values for these days are presented in Table \ref{tb1}, with most exceeding 80, and only one day falling below 60.
The average rank in 2024 is 89.8, higher than the 77.1 in 2015, suggesting that the registration performance of the APRIL method is influenced by data quality. Nevertheless, the precision remained robust in most cases, consistently within 0.2 pixels ($\sim 0.4''$), and could achieve an accuracy better than 0.05 pixel ($\sim0.1''$) under optimal conditions. The minimum rank of 25 for September 30, 2015, was primarily due to cloudy weather and dust accumulation on the objective lens, which introduced significant noise into the coronal data.

\begin{deluxetable*}{cccccccccc}
%\digitalasset
\tablewidth{0pt}
\tablecaption{Summary table of the registration results.}\label{tb1}
\tablehead{
\colhead{Date} & \colhead{Sampled Counts} & \colhead{Image Size} & \colhead{Convergence Rate} & \colhead{$\overline{X_{std}}$} &\colhead{$\overline{Y_{std}}$} &\colhead{$\overline{R_{std}}$} &\colhead{$\overline{\theta_{std}}\ (10^{-5})$} & \colhead{$\overline{R_{\odot}}$} & \colhead{Rank}
}
\startdata
20150104 & 50 / 50 & (1024, 1024) & 100.0\% & 0.137 & 0.098 & 0.058 & 5.23 & 308.758 & 83.3 \\
20150113 & 46 / 46 & (1024, 1024) & 100.0\% & 0.128 & 0.105 & 0.060 & 6.03 & 308.196 & 82.9 \\
20150202 & 40 / 40 & (1024, 1024) & 100.0\% & 0.224 & 0.178 & 0.097 & 6.42 & 307.610 & 71.2 \\
20150215 & 19 / 19 & (1024, 1024) & 100.0\% & 0.248 & 0.212 & 0.109 & 7.61 & 307.107 & 67.1 \\
20150302 & 33 / 33 & (1024, 1024) & 93.8\% & 0.229 & 0.204 & 0.121 & 3.51 & 306.589 & 65.5 \\
20150308 & 44 / 44 & (1024, 1024) & 95.3\% & 0.136 & 0.118 & 0.053 & 5.46 & 305.661 & 81.3 \\
20150504 & 9 / 9 & (1024, 1024) & 88.9\% & 0.064 & 0.062 & 0.029 & 3.60 & 300.831 & 88.1 \\
20150514 & 11 / 11 & (1024, 1024) & 100.0\% & 0.065 & 0.058 & 0.021 & 3.26 & 299.885 & 91.8 \\
20150711 & 16 / 16 & (1024, 1024) & 100.0\% & 0.117 & 0.087 & 0.043 & 5.84 & 298.858 & 85.7 \\
20150930 & 15 / 15 & (1024, 1024) & 100.0\% & 0.621 & 0.433 & 0.246 & 24.77 & 303.124 & 25.0 \\
20151027 & 4 / 4 & (1024, 1024) & 100.0\% & 0.107 & 0.076 & 0.034 & 6.03 & 305.676 & 87.9 \\
20151106 & 10 / 10 & (1024, 1024) & 100.0\% & 0.197 & 0.085 & 0.069 & 3.67 & 306.268 & 80.3 \\
20151123 & 38 / 38 & (1024, 1024) & 100.0\% & 0.128 & 0.097 & 0.053 & 6.03 & 307.435 & 84.1 \\
20151223 & 37 / 37 & (1024, 1024) & 91.4\% & 0.083 & 0.082 & 0.039 & 4.01 & 308.356 & 86.1 \\
20151226 & 52 / 52 & (1024, 1024) & 92.1\% & 0.157 & 0.143 & 0.074 & 6.77 & 308.300 & 76.3 \\
\hline
20240101 & 261 / 784 & (1400, 1400) & 100.0\% & 0.022 & 0.020 & 0.008 & 2.05 & 371.422 & 97.7 \\
20240118 & 298 / 895 & (1400, 1400) & 99.4\% & 0.081 & 0.103 & 0.040 & 3.62 & 369.877 & 88.9 \\
20240209 & 286 / 765 & (1400, 1400) & 99.6\% & 0.106 & 0.164 & 0.047 & 5.62 & 370.141 & 84.8 \\
20240218 & 292 / 860 & (1400, 1400) & 100.0\% & 0.027 & 0.031 & 0.008 & 2.57 & 369.921 & 97.0 \\
20240306 & 267 / 801 & (1400, 1400) & 97.9\% & 0.176 & 0.165 & 0.072 & 8.38 & 365.912 & 79.6 \\
20240323 & 199 / 598 & (1400, 1400) & 95.6\% & 0.156 & 0.109 & 0.077 & 6.20 & 364.693 & 82.0 \\
20240412 & 77 / 231 & (1400, 1400) & 100.0\% & 0.028 & 0.027 & 0.019 & 3.41 & 363.753 & 96.1 \\
20240414 & 288 / 864 & (1400, 1400) & 89.8\% & 0.128 & 0.097 & 0.094 & 6.86 & 362.861 & 80.5 \\
20240514 & 136 / 410 & (1400, 1400) & 100.0\% & 0.038 & 0.051 & 0.015 & 3.25 & 361.849 & 94.9 \\
20240516 & 8 / 24 & (1400, 1400) & 100.0\% & 0.022 & 0.027 & 0.009 & 2.11 & 361.038 & 97.1 \\
20240603 & 15 / 45 & (1400, 1400) & 100.0\% & 0.028 & 0.027 & 0.015 & 2.97 & 359.640 & 96.4 \\
20240919 & 279 / 839 & (1400, 1400) & 89.5\% & 0.191 & 0.175 & 0.113 & 5.21 & 361.528 & 72.8 \\
20240923 & 94 / 284 & (1400, 1400) & 92.1\% & 0.216 & 0.106 & 0.089 & 10.71 & 362.708 & 78.1 \\
20241014 & 121 / 365 & (1400, 1400) & 95.6\% & 0.108 & 0.103 & 0.052 & 5.14 & 364.254 & 85.9 \\
20241031 & 66 / 200 & (1400, 1400) & 100.0\% & 0.025 & 0.036 & 0.011 & 2.01 & 367.733 & 96.5 \\
20241111 & 448 / 1346 & (1400, 1400) & 99.5\% & 0.029 & 0.042 & 0.014 & 2.05 & 368.028 & 95.7 \\
20241125 & 194 / 584 & (1400, 1400) & 100.0\% & 0.028 & 0.034 & 0.010 & 2.13 & 369.021 & 96.6 \\
20241204 & 447 / 1341 & (1400, 1400) & 95.2\% & 0.026 & 0.053 & 0.014 & 2.49 & 369.947 & 94.3 \\
20241214 & 389 / 1168 & (1400, 1400) & 91.8\% & 0.059 & 0.065 & 0.027 & 3.62 & 370.393 & 90.7 \\
\enddata
%\tablecomments{Table 2 is published in its entirety in the electronic 
%edition of the {\it Astrophysical Journal}.  A portion is shown here 
%for guidance regarding its form and content. The {\tt\string \digitalasset}\ command highlights the Table title to visually indicate to the reader that there is data associated with this table.}
\end{deluxetable*}

%Due to space constraints, Table \ref{tb1} presents only the daily average values for select parameters. For detailed data, we have assembled a comprehensive catalog for each date, detailing the results of the APRIL method on each coronal image, including the similarity transformation parameters, solar radius, position and radius of the occulter, and more. These catalogs are available at \url{https://nadc.china-vo.org/res/file_upload/download?id=48628}. 

\begin{figure*}[ht!]
	\plotone{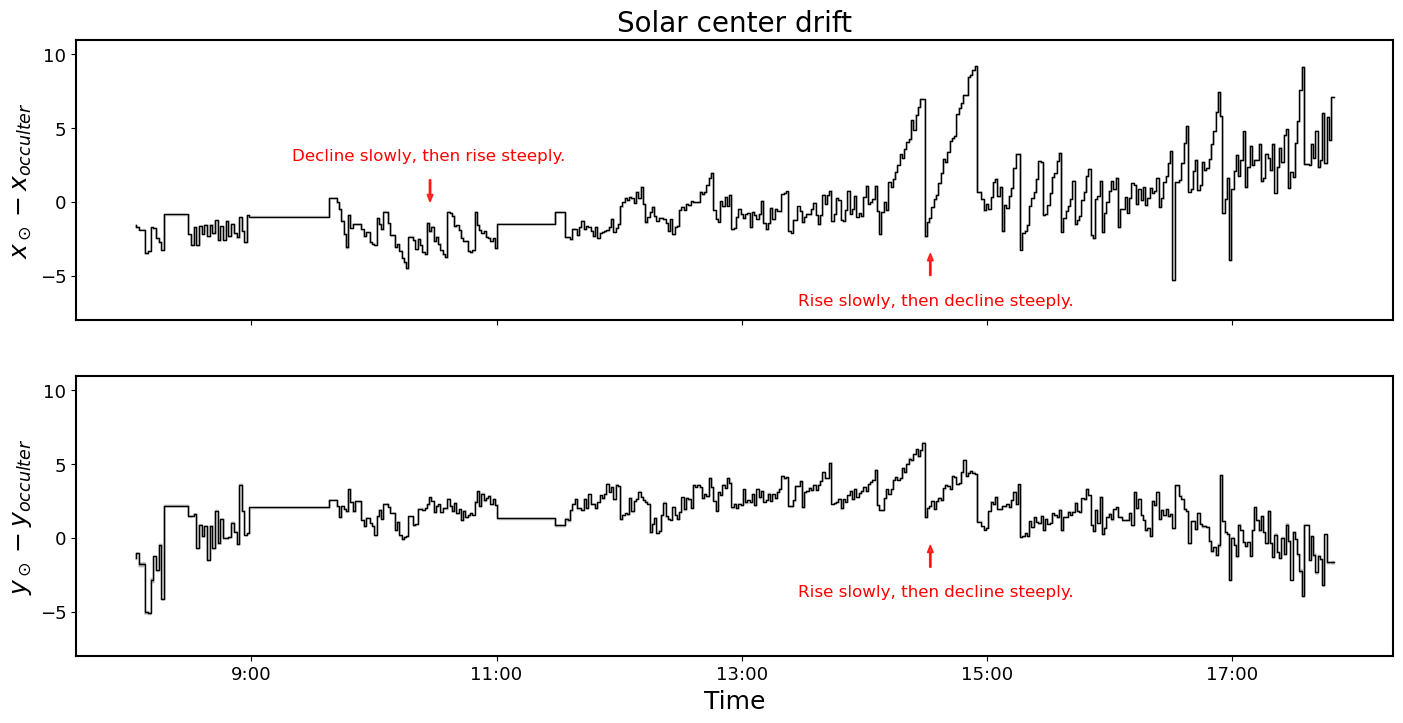}  % 控制宽度为页面宽度
	\caption{Drift of YOGIS solar center relative to the occulter on November 11, 2024, calculated using the APRIL method (unit: pixel). The upper and lower panels illustrate the drift in the horizontal and vertical directions, respectively. The black solid line and the surrounding gray area represent the drift and the standard deviation. The error can hardly be seen since it is too small. \label{fig:drift}}
\end{figure*}

Still taking November 11, 2024 as an example, the drift of the solar center relative to the occulter is shown in Figure \ref{fig:drift}. The drift exhibits a distinct pattern, first moving slowly in one direction and then rapidly returning, as shown in the red arrows of Figure \ref{fig:drift}. The slow drifting phase is caused by errors in the guiding system, while the return phase results from the observer's manual adjustment of the telescope pointing. Based on these data, we can analyze the errors in the guiding system and improve its accuracy in the future.

Finally, 100 days of coronal data spanning from 2013 to 2024 are selected and mapped to HCC by the APRIL method. The registration parameters of each image are recorded in a catalog, with one catalog corresponding to each observation day, accessible at PaperData (doi: \href{https://doi.org/10.12149/101571}{10.12149/101571}, see the Catalogs.zip archive) %\url{https://nadc.china-vo.org/res/file_upload/download?id=51720}. 
%day, accessible at PaperData (doi: \href{https://doi.org/10.12149/101571/}{10.12149/101571}).
%\url{https://nadc.china-vo.org/res/file_upload/download?id=51720}. 
The registered images are saved as FITS files,
%All coronal green-line images registered using the APRIL method are saved as FITS files, 
with the Sun centered in the image and the solar north pole aligned vertically upwards.
%rotated to the top.
 Critical HCC information and the occulter position are recorded in the header. These data can easily be converted to HCC maps, either through the Sunpy package in Python or the Solar SoftWare (SSW) library in IDL. The FITS files for the given dates are accessible at PaperData (doi: \href{https://doi.org/10.12149/101571}{10.12149/101571}). Note that these FITS files are currently Level 0.5 data, as the coronal emission intensity has not been calibrated yet. We are actively working on it and plan to release Level 1 coronal data in the future. 
The images of each day are compiled into a movie for quick look, which can be accessed at PaperData (doi: \href{https://doi.org/10.12149/101571}{10.12149/101571}, see the Movies.zip archive).

\section{Conclusion}
In this study, we present a coronal image registration method, APRIL, by which coronal green-line images and SDO/AIA 211 \r{A} images are automatically registered with high precision, enabling accurate mapping of coronal images to the HCC. 
The APRIL method is validated using a large amount of coronal data, including the coronal green-line images observed by the two coronagraphs, YOGIS and SICG, at Lijiang Observatory, Yunnan Observatories. It achieves a registration accuracy better than $0.1''$ under optimal data quality, while maintaining a precision no worse than $0.4''$ in most cases. The results highlight the method's robustness in processing coronal green-line data across varying observation dates and different coronagraphs. It will soon be applied to the upcoming 20-cm coronagraph of the China Meteorological Administration, which has been fully developed and is ready for installation. Additionally, it will also be used for the large 60-cm coronagraph currently under development.

In summary, this registration method significantly improves the mapping precision of coronal images, particularly for coronagraphs lacking guiding systems, such as the recently developed 50-mm balloon-borne coronagraph \citep{Liu21,Lin23b}. The registered images enable more accurate studies of coronal transient activities and support high-precision localization of small regions in multi-device and multi-waveband observations. Furthermore, the method eliminates the need to average multiple frames beforehand. This facilitates the detection of short-timescale coronal activities and high-frequency oscillations using ground-based coronagraphs.

It should be noted that the APRIL method requires the presence of clear coronal structures around the solar disk as a prerequisite. As a result, it is not applicable for coronal images obtained during solar minimum.

In the UCoMP Data User's guide\footnote{\url{https://mlso.hao.ucar.edu/files/UCoMP-Data-User's-Guide-for-Distrubtion_Web_2023nov29.pdf}}, researchers note that they are evaluating various image alignment methods for UCoMP data. 
In the foreseeable future, the APRIL method could be applied to register UCoMP green-line historical data with 211 \r{A} images, enabling not only the alignment of green-line images over an observing day but also the derivation of an accurate HCC.
However, the current method is not suitable for data observed in other wavebands. We plan to enhance the method to enable registration between images observed by the same device, making it applicable to all wavebands.

In the future, we plan to apply the APRIL method to correct errors in the YOGIS guiding system, ensuring that the occulter and solar disk remain concentric at all times. This will help minimize stray light levels and enhance data consistency.

%% Please use the acknowledgment and contribution environments. This will 
%% be anonomyized when the "anonymous" style option is used. 
\begin{acknowledgments}
We sincerely thank the NAOJ team for the long-term 10-cm coronagraph collaborations. We acknowledge the data resources from the National Space Science Data Center, National Science \& Technology Infrastructure of China (\href{http://www.nssdc.ac.cn}{http://www.nssdc.ac.cn}). We acknowledge the use of data from the Chinese Meridian Project. This study is supported by the National Natural Science Foundation of China (NSFC 12373063, 11533009, 12173086, 12163004) and the Sichuan Science and Technology Program (2025ZNSFSC0877)
\end{acknowledgments}

\begin{contribution}
%%This section gives authors the space to recognize author contributions. The text inside this environment is NOT counted towards the total word quanta. At a minimum, manuscripts are expected to include this text:

All authors contributed equally to this work.

%% But authors are expected to provide more specific details, e.g. 
%%
%%SC was responsible for writing and submitting the manuscript.
%%WWM came up with the initial research concept and edited the manuscript.
%%OTS obtained the funding and edited the manuscript.
%%EBF provided the formal analysis and validation. He also edited the manuscript.
%%GEH Supervised the undergraduates, wrote the software and administers the project github and Zenodo repositories.
%%
%% Authors can use the Contributor Role Taxonomy (CRediT) at
%% for ideas on how write a good statement tailored to their needs.

\end{contribution}

%\facilities{HST(STIS), Swift(XRT and UVOT), AAVSO, CTIO:1.3m, CTIO:1.5m, CXO}
\facilities{YOGIS, SICG, SDO(AIA)}

\software{astropy \citep{Astropy},  
          sunpy \citep{sunpy}
          }


%% For this sample we use BibTeX plus aasjournalv7.bst to generate the

\bibliography{reference}{}
\bibliographystyle{aasjournalv7}



\end{document}